\documentclass [12pt] {article}
\usepackage{amsmath}
\usepackage{amsfonts}
\newcommand{\ed}{\end{document}}

\begin{document}
\title{Gauge algebra of irreducible theories in 
the Sp(2)-symmetric BRST formalism} 
\author{A.~V.~Bratchikov \\ Kuban State
Technological University,\\ Krasnodar, 350072,
Russia
} 
\date {May,\,2012} 
\maketitle

\begin{abstract}
An explicit solution to classical master equations of the Sp(2)-symmetric Hamiltonian BRST
quantization scheme is presented in the case of irreducible gauge theories. A realization of the observable algebra is constructed.
\end {abstract}



\section{Introduction}

Let M be a phase space with the phase coordinates $p_i, q^i$ $ i=1,\ldots, n,$ and the Poisson bracket $\{.\,,.\}.$ Let $T_\alpha=T_\alpha(p, q), \alpha =1,\ldots,m,$ $m<n,$ be first class constraints of a Hamiltonian system,
\begin{eqnarray*}
\{T_\alpha, T_\beta\}
\approx 0.
\end{eqnarray*}
The weak equality $\approx$ means equality on the constraint surface 
\begin{eqnarray*}\Sigma:\qquad T_{\alpha}=0.
\end{eqnarray*}
Let $P$ denote the Poisson algebra of first class functions,
\begin{eqnarray*}
P =\{f(p, q)\,|\, 
\{f,T_\alpha \}\approx 0
\}
,
\end{eqnarray*}
and let 
\begin{eqnarray*}
J =\{u(p, q)\,|\, u\approx 0\}.
\end{eqnarray*}
Elements of the Poisson algebra $P/J$ are called classical observables. The Hamiltonian $H_0(p, q)$ is assumed to be a first class function. These definitions correspond to the Dirac quantization without gauge fixing \cite {D1}.

There are different realizations of $P/J.$ In the gauge fixing method \cite {D2} auxiliary constraints are introduced and the original Poisson algebra is replaced by the Dirac one.
The algebra $P/J$ is isomorphic to a quotient Dirac bracket algebra.
Some other realizations of the observable algebra without extending of the original phase space were given in Refs.  
\cite{Sn}, \cite{Br1}.   

In the hamiltonian BRST theory 
with the BRST charge $\Omega$ 
the realization of $P/J$ looks like $U/V$ where $U$ is the space of solutions to the equation
\begin{eqnarray*}
\{\Omega,\Phi\}=0
\end{eqnarray*}
with certain boundary conditions \cite {FHST}.
 
In the present paper we study the observable algebra 
of irreducible gauge theories in the framework of the Sp(2)-symmetric BRST formalism. 
The extended BRST symmetry was discovered in Refs. \cite {CF}, \cite{O}. 
In the Hamiltonian formalism it is generated by the charges $\Omega^a,$ $a=1,2,$ satisfying the master equations 
\begin{eqnarray}
\label {3}
\{\Omega^a,\Omega^b\}=0.
\end{eqnarray}
Observables are determined by solutions to the equation
\begin{eqnarray}
\label {4}
\{\Omega^a,\Phi\}=0.
\end{eqnarray}
A solution to the generating equations (\ref{3}), (\ref{4}) for rang-1 theories was found in Ref. \cite{Sp}.
In the general case an algorithm for the construction of a solution to these equations was given in Refs. \cite {BLT}, \cite {BLT2} 
(see also \cite {GH}).
However, the problem of finding 
the charges $\Omega^a$ and observables has not been solved.
The goal of this paper is to present an explicit solution to Eqs. (\ref{3}), (\ref{4}) and construct a realization of $P/J.$

The paper is organized as follows.
In section 2, we review the master equations of the Sp(2)- symmetric Hamiltonian BRST theory and introduce notations. 
An explicit formula  
for $\Omega^a$ is given in section~3. 
The realization of the observable algebra is described in section 4.

In what follows the Grassmann parity and new ghost number 
of a function $X$ are denoted by $\epsilon (X)$ and $\mbox{ngh}(X),$ 
respectively.
The constraints are supposed to be of definite Grassmann parity $\epsilon(T_\alpha)=\epsilon_\alpha.$ For a function $X^{a_1a_2\ldots a_n}$
\begin{eqnarray*}X^{\{a_1a_2\ldots a_n\}}=X^{a_1a_2\ldots a_n}+ 
\mbox {cycl. perm.}\, 
({a_1,a_2,\ldots, a_n}).
\end{eqnarray*}

\section {Master equations 
}
An extended phase space of the theory 
under consideration 
is parametrized   
by the canonical variables 
\begin{eqnarray*}
{\cal G}= (p_i,q^i;{\cal P}_{\alpha  a},C^{\alpha  a};\lambda_\alpha,\pi^\alpha),
\end{eqnarray*} 
\begin{eqnarray*}\epsilon(p_i)=\epsilon(q^i)=\epsilon_i,\qquad 
\epsilon({\cal P}^{\alpha a})=\epsilon(C^{\alpha a})=\epsilon_\alpha+1,\qquad 
\epsilon(\lambda_\alpha)=\epsilon(\pi^\alpha)=\epsilon_\alpha,
\end{eqnarray*} 
\begin{eqnarray*}
\mbox {ngh}(p_i)=
\mbox {ngh}(q^i)=0, \qquad \mbox {ngh}({\cal P}_{\alpha a})=-1,\qquad \mbox {ngh}(C^{\alpha a})=1, 
\end{eqnarray*} 
\begin{eqnarray}
\label{100}
\mbox {ngh}(\pi^\alpha)=2,\qquad \mbox {ngh}(\lambda_\alpha)=-2. 
\end{eqnarray} 
The Poisson bracket is given by
\begin{eqnarray*}
\{X, Y\}= 
\frac {\partial X} {\partial q^i} \frac 
{\partial Y} {\partial p_i}+ 
\frac {\partial X} {\partial C^{\alpha a}} \frac 
{\partial Y} {\partial {\cal P}_{\alpha a}}+
\frac {\partial X} {\partial \pi^{\alpha}} \frac 
{\partial Y} {\partial \lambda_\alpha} - (-1)^{\epsilon (X)\epsilon (Y)}(X\leftrightarrow Y).
\end{eqnarray*}  
Derivatives with respect to the generalized momenta $p,$ ${\cal P},$ 
$\lambda$ are understood as left-hand, and those with respect to the generalized coordinates $q,$ $C,$ $\pi$ as right-hand ones.

We assume that $T_\alpha$ are independent and satisfy the regularity conditions. This means that there are some functions $F_{\alpha'}(p,q),$ 
such that $(F_{\alpha'},T_{\alpha})$  can be locally taken as new coordinates in the original phase space.
This assumption allows changing variables:
$(p,q)\to
\xi=(\xi_{\alpha}, \xi_{\alpha'}),$
\begin{eqnarray*} 
\xi_{\alpha} = T_{\alpha}(p,q),\qquad \xi_{\alpha'} =  F_{\alpha'}(p,q).
\end{eqnarray*}
In what follows we use only the phase variables $(\xi,{\cal P},C,\lambda,\pi).$
The constraint surface $\Sigma$ looks like
\begin{eqnarray*}\xi_{\alpha}=0.
\end{eqnarray*}
The Poisson bracket is denoted by $\{.\,,.\}'.$ 

With respect to the new variables Eq. (\ref{3}) takes the form
\begin{eqnarray} \label{oa1}
\{\Omega^{\prime a},\Omega^{\prime b}\}'=0,
\end{eqnarray}
where $\Omega^{\prime a}(\xi,{\cal P},C,\lambda,\pi)=\Omega^{a}(p,q,{\cal P},C,\lambda,\pi).$
The charges $\Omega^{\prime a}$ also satisfy the 
conditions
\begin{eqnarray*}
\epsilon (\Omega^{\prime a})=1, \qquad \mbox {ngh}(\Omega^{\prime a})=1,
\end{eqnarray*}
\begin{eqnarray} 
\label{o1}
\left.
\frac 
{\partial \Omega^{\prime a}} 
{\partial C^{\alpha b}}\right|_{C=\pi={\cal P}=\lambda=0} =\xi_\alpha \delta^a_b,
\qquad \left.\frac {\partial \Omega^{\prime a}} {\partial {\pi^{\alpha}}}\right|_{C=\pi=\lambda=0} =\varepsilon^{ab} {\cal P}_{\alpha b},
\end{eqnarray}
where 
$$
\varepsilon^{ab}=
\left(
\begin{array} {cc}
0&1\\
-1&0
\end{array}  
\right).
$$
One can write
\begin{eqnarray} \label{us2}
\Omega^{\prime a}= \Omega_1^a+\Pi^a,
\end{eqnarray}
where
\begin{eqnarray}
\Omega_1^a=\xi_\alpha C^{\alpha a}
+ \varepsilon^{ab} {\cal P}_{\alpha b}\pi^\alpha,  \qquad\Pi^a= \sum_{n\geq 2} \Omega^a_n,\qquad  
\Omega_n^a \sim  C^{n-m}\pi^m.
\end{eqnarray}

Let $
N$ be the counting operator 
\begin{eqnarray*} 
N = \xi_\alpha \frac { \partial_l} {\partial {\xi}_\alpha}+{\cal P}_{\alpha a} \frac {\partial} {\partial {\cal P}_{\alpha a}}+
\lambda_\alpha \frac {\partial}  {\partial \lambda_\alpha},
\end{eqnarray*}
and let 
\begin{eqnarray*} 
W^a=\xi_\alpha  \frac {\partial }  {\partial  {\cal P}_{\alpha a}} + \varepsilon^{ab} {\cal P}_{\alpha b}\frac {\partial } {\partial \lambda_\alpha}
+ (-1)^{\epsilon_\alpha}\varepsilon^{ab}\pi^\alpha \frac {\partial_l }  {\partial C^{\alpha  b}},  
\end{eqnarray*}
\begin{eqnarray*} 
\Gamma_a = {\cal P}_{\alpha a} \frac {\partial_l } {\partial \xi_\alpha  } - \varepsilon_{ab} \lambda_\alpha
 \frac {\partial } {\partial  {\cal P}_{\alpha b}} ,\qquad M=\Gamma_aW^a,  
\end{eqnarray*}
where $\varepsilon^{ab}\varepsilon_{bc}=\delta^{a}_c.$
Then 
\begin{eqnarray*}
W^{\{a}W^{b\}}=0,  \qquad \Gamma_{\{a}\Gamma_{b\}}=0,\qquad W^a\Gamma_b+\Gamma_bW^a=\delta^a_bN,
\end{eqnarray*}
\begin{eqnarray*} 
NW^a=W^aN , \qquad N\Gamma_a=\Gamma_a N,
\end{eqnarray*}
\begin{eqnarray*} 
M^2W^a=NMW^a,\qquad
\Gamma_a M^2= N\Gamma_a M,
\end{eqnarray*}
\begin{eqnarray} 
\label{us4} 
M^n=(2^{n-1}-1)N^{n-2}M^2-(2^{n-1}-2)N^{n-1}M, \qquad n\geq 3. 
\end{eqnarray}

Substituting (\ref {us2}) in (\ref {oa1}) we get
\begin{eqnarray}
\label{dur}
W^{\{a}\Pi^{b\}}+F^{ab}+ A^{\{a}\Pi^{b\}}+ \{\Pi^a,\Pi^b\}'  =0,
\end{eqnarray}
where \begin{eqnarray*} 
F^{a b} =C^{\alpha a}\{\xi_\alpha, \xi_\beta\}' C^{\beta b},\qquad 
A^a=C^{\alpha a}\{\xi_\alpha,\,.\, \}'.
\end{eqnarray*}

\section {Solving the master equations}

Let ${\cal G}'$ be the set of variables $\xi,{\cal P},C,\lambda,\pi,$ and let 
${\cal V}$ be the space of the formal power series in the variables ${\cal G}'$ which
vanish on $\Sigma$ at ${\cal P}=\lambda=0.$
The space ${\cal V}$ splits as 
\begin{eqnarray*} 
{\cal V}= \bigoplus_{m\geq 1} {{\cal V}}_m. 
\end{eqnarray*}
with  $
N X=mX$ for $X\in {\cal V}_m.$
It is clear that the operator ${N:{\cal V} \to {\cal V}}$ is invertible. 
Let ${S}^n,$ $n \geq 1,$ denote the space of the functions $X^{a_1\ldots a_n}\in {\cal V}$ which are symmetric under permutation of any indices. Equations (\ref{100}) imply  that $\Omega^{\prime a} \in {S}^1,$ and  $\{\Omega^{\prime a},\Omega^{\prime b}\}'\in {S}^2.$

Define the operators $
W: S^n\to S^{n+1},$ and $\Gamma: S^{n+1}\to S^{n}, $
as 
\begin{eqnarray*} 
(WX)^a=W^aX, \qquad (\Gamma X)=\Gamma_{a}X^{a},\qquad n=0,
\end{eqnarray*} 
\begin{eqnarray*} \label {or}
 (WX)^{a_1\ldots a_{n+1}}= W^{\{a_1}X^{a_2\ldots a_{n+1}\}},\qquad 
(\Gamma X)^{a_1\ldots a_{n}}=\Gamma_{a}X^{a_1\ldots a_{n} a},\qquad n\geq 1,
\end{eqnarray*} 
where $S^{0}={\cal V}.$ For $X\in S^0$ we set $$\Gamma X=0.$$
One can directly verify that
\begin{eqnarray*} 
W^2=0,\qquad \Gamma^2=0, \qquad  \Gamma M=(M-N)\Gamma,
\end{eqnarray*} 
\begin{eqnarray} \label {oor}
WM=(M+N)W, \qquad (\Gamma W+W\Gamma)X=(nN+M)X, 
\end{eqnarray} 
where $X\in S^n,$ $n \geq 0.$ 
By using (\ref{us4}), we get  
\begin{eqnarray*} 
(nN+M)^{-1}=\frac 1 n N^{-1}-\frac {1} {n(n+2)(n+1)}((n+3)MN^{-2}-  M^{2}N^{-3}),\quad n \geq 1.
\end{eqnarray*}

Let $Q: S^{n}\to S^{n}$ 
be defined by 
\begin{eqnarray*}QX =\frac 1 6 ({11}N^{-1}- 6 MN^{-2}+ M^2 N^{-3})X,\qquad X \in S^0,
\end{eqnarray*}
\begin{eqnarray*}QX =(nN+M)^{-1}X,\qquad X \in S^{n},\qquad n\geq 1.\,\,\,\,\,\,\,\,\,\,\,\,\,\,\,\,\,\,\,
\end{eqnarray*}
Then $W^{+}=Q\Gamma $ is a generalized inverse of $W$
\begin{eqnarray} \label {or7}
WW^+W=W. 
\end{eqnarray} 
From  (\ref{oor}) it follows that \begin{eqnarray*} 
(W^+)^2=0, 
\end{eqnarray*} 
and for any $X\in S^n,$ $n \geq 1,$
\begin{eqnarray} \label {dor}
X= W^+WX + WVW^+X. 
\end{eqnarray} 
Here 
\begin{eqnarray*} 
V=\frac {1} {n(n+2)(n+1)}\left(n(n^2+4n+6)I-(n-4)MN^{-1}-2M^2N^{-2}\right),
\end{eqnarray*}
and $I$ is the identity map.

For any $X\in S^1$, $Y\in S^n,$ $n\geq 1,$ we define the bracket ${[.\,,.]:S^1\times S^n\to S^{n+1}}$ as
\begin{eqnarray*}[X,Y]^{aa_1,\ldots a_n}=
\{X^{\{a},Y^{a_1,\ldots a_n\}}\}'.
\end{eqnarray*}
Then (\ref{dur}) can be written as
\begin{eqnarray}
\label{durs}
W\Pi+F+ A\Pi+ [\Pi,\Pi]  =0,
\end{eqnarray}
where
\begin{eqnarray*}
\Pi=(\Pi^1,\Pi^2), \qquad F=(F^{ab}), 
\qquad (A\Pi)^{ab}=
A^{\{a}\Pi^{b\}}.
\end{eqnarray*}

Applying 
the operator $WW^+$ to (\ref {durs}), we get 
 \begin{eqnarray}
\label{d3}
W\Pi+WW^+(F+ A\Pi+ [\Pi,\Pi]) =0.
\end{eqnarray}
From (\ref{d3}) it follows that 
\begin{eqnarray}
\label{seo} \Pi=\Upsilon-W^+(F+ A\Pi+ [\Pi,\Pi]),
\end{eqnarray}
where 
\begin{eqnarray*} 
\Upsilon\in S^1, \qquad W \Upsilon=0, \qquad
\Upsilon= \sum_{n\geq 2}\Upsilon^{(n)},\qquad  
\Upsilon^{(n)} \sim  C^{n-m}\pi^m.
\end{eqnarray*}
If $\Pi$ is a solution to (\ref{seo}) then
\begin{eqnarray}\label{see}
W^+\Pi= W^+\Upsilon.
\end{eqnarray}

Let $\langle .\,,. \rangle:  S^1\times S^1 \to S^1$  be defined by
\begin{eqnarray} \label {obosr}
\langle X_1,X_2 \rangle = -\frac 1 4 (I+W^+A)^{-1}W^+\left([ X_1,X_2]+[X_2,X_1] \right),
\end{eqnarray} 
where  
\begin{eqnarray*} (I+W^+A)^{-1}=\sum_{m\geq 0}(-1)^m(W^+A)^m. 
\end{eqnarray*}
Then we have
\begin{eqnarray}
\label{dursd}
\Pi= \Pi_0+\frac 1 2 \langle \Pi,\Pi \rangle,
\end{eqnarray}
where 
\begin{eqnarray*}
\Pi_0 = (I+W^+A)^{-1}\left(\Upsilon - W^+F\right).
\end{eqnarray*} 

Let us show that a solution to (\ref{dursd}) satisfies (\ref {durs}). We shall use the approach of ref. \cite {F}.
The Jacobi identities
\begin{eqnarray*} 
\{\Omega^{\prime a},\{\Omega^{\prime b},\Omega^{\prime c}\}'\}'+\mbox {cycl. perm.}\, (a,b,c)=0
\end{eqnarray*}
imply 
\begin{eqnarray} \label{o2}
[\Omega,G]=0,
\end{eqnarray}
where $\Omega=(\Omega^{\prime a})$ and $G$ is left-hand side of (\ref{durs}),   
\begin{eqnarray*}
G= W\Pi+F+ A\Pi+ [\Pi,\Pi]. 
\end{eqnarray*}
Equation (\ref{o2}) can be written as 
\begin{eqnarray} \label{aux}
WG+AG +[\Pi,G]=0.
\end{eqnarray}
Consider (\ref {aux}), where $\Pi$ is a solution to (\ref{seo}), with the boundary 
condition 
\begin{eqnarray}
\label{au23}
W^+G=0.
\end{eqnarray}
Applying $W^+$ to (\ref {aux}), and using (\ref {au23}), we get
\begin{eqnarray*}
G+W^+(AG +[\Pi,G])=0.
\end{eqnarray*}
From this by iterations it follows that $G=0.$

For checking (\ref{au23}) we have 
\begin{eqnarray*}
W^+G=W^+W\Pi+W^+(F+ A\Pi+ [\Pi,\Pi])=W^+W\Pi+\Upsilon-\Pi,
\end{eqnarray*}
and therefore by (\ref{dor}) and (\ref{see}), $W^+G=0.$

To obtain an explicit expression for $\Pi$ we introduce the functions 
\begin{eqnarray*}\langle \ldots.  \rangle:
 \left(S^1\right)^m 
 \to  {S^1},\qquad m=1,2,\ldots,
  \end{eqnarray*}
which recursively 
defined by $\langle X \rangle=X,$ 
\begin{eqnarray*} \label {u}
\langle X_1,\ldots ,X_m \rangle =
\frac 1 2 \sum_{r=1}^{m-1} \sum_{1\leq i_1<\ldots < i_r \leq m } \langle \langle X_{i_1},\ldots,X_{i_r} \rangle,
\langle X_1,\ldots,\widehat X_{i_1},\ldots,\widehat X_{i_r},\ldots,X_{m} \rangle 
 \rangle
\end{eqnarray*} 
if $m=3,4,\ldots,$ where $\widehat{X}$ means that ${X}$ is omitted. Recall that $\langle X_1,X_2 \rangle$ is defined in (\ref {obosr}). 
Using induction on $m$ one easily verifies that $ \langle X_1,\ldots ,X_m \rangle$ is an $m-$linear symmetric function.

For $m\geq 2,1\leq i,j\leq m,$ let  
\begin {eqnarray*}P^m_{ij}:  (S^1)^m\to (S^1)^{m-1}
\end{eqnarray*}   
be defined by
\begin{eqnarray*} 
P^m_{ij}( X_1,\ldots ,X_m ) = 
( \langle X_i,{X}_{j}\rangle ,X_1 ,\ldots,\widehat{X}_{i},\ldots,\widehat{X}_{j},\ldots,X_{m} ). 
\end{eqnarray*} 
If $X \in S^1$ is given by  
\begin{eqnarray} 
 X = P^2_{12}P^3_{
i_{m-2}j_{m-2}
}\ldots P^{m-1}_
{i_2j_2}
P^m_{i_1j_1}
(X_1,\ldots ,X_m )
\label {ku}
\end{eqnarray} 
for some $ (i_1j_1),\ldots,(i_{m-2}j_{m-2}),$ we say that $X$ is a descendant of $(X_1,\ldots ,X_m ).
$
A descendant of $X\in S^1$ is defined as $X.$
One can show that $\langle X_1,\ldots,X_m \rangle$ equals the sum of all the descendants of  $(X_1,\ldots ,X_m )$ \cite {Br2}.
\vspace{1mm}
For example, \begin{eqnarray*} \langle X_1,X_2,X_{3}\rangle=
\langle \langle X_{1},X_{2}\rangle, X_3 \rangle+\langle \langle X_{1},X_{3}\rangle, X_2 \rangle+\langle  \langle X_2, X_3 \rangle, X_{1} \rangle. 
\end{eqnarray*} 
This remembers Wick's theorem in QFT.

We can now write the general solution to eq. (\ref {dursd}). 
It is given 
by \cite{Br2}
\begin{eqnarray*} 
\Pi= \langle e^{\Pi_0} \rangle,
\end{eqnarray*}
where
\begin {eqnarray*} \langle e^{\Pi_0} \rangle =\sum_{m\geq 0} \frac {1} {m!}\langle \Pi^m_0 \rangle 
, \qquad \langle \Pi^0 \rangle=0,
\qquad \langle \Pi_0^n \rangle= \langle \underbrace{\Pi_0,\ldots ,\Pi_0}_{n 
\text\,\, {\rm times}
}\rangle,\qquad n \geq 1.
\end{eqnarray*} 
Finally, we get
\begin{eqnarray*} \label {orrod}
\Omega = \Omega_1+ \langle e^{\Pi_0} \rangle, 
\end{eqnarray*}
where $\Omega_1=(\Omega_1^a).$


\section {Realization of observables 
} 
With respect to 
the variables $(\xi,{\cal P},C,\lambda,\pi)$ Eq. (\ref{4}) takes the form
\begin{eqnarray}
\label{pps}
\{\Omega^{\prime a},\Phi'\}'=0,
\end{eqnarray}
where $\Phi'(\xi,{\cal P},C, \lambda,\pi)=\Phi(p,q,{\cal P},C, \lambda,\pi).$
The boundary conditions read
\begin{eqnarray}
\label{pps2}
\mbox{\rm ngh} (\Phi')=0,\qquad \left.\Phi'\right|_{C=\pi=0}=\Phi_0
,\qquad 
\bar \Gamma(\Phi'-\Phi_0)=0,
\end{eqnarray}
where $\Phi_0(\xi) \in P,$ $\bar \Gamma=\epsilon^{ab}\Gamma_a\Gamma_b.$ 

The function $\Phi'$ can be written as  
\begin{eqnarray} 
\label{us0} 
\Phi' = \Phi_0+K,\qquad K=\sum_{n\geq 1}\Phi^{(n)},\qquad 
\Phi^{(n)}\sim C^{n-m}\pi^m.
\end{eqnarray}
Substituting (\ref{us0}) in (\ref{pps}), we get  
\begin {eqnarray}
\label{qsp}
W^a K+ \{\Omega^{\prime a}, \Phi_0\}'+ A^a K + \{\Pi^{a},K\}'=0, 
\end {eqnarray}
For $X\in S^1$, $Y\in S^0$ denote 
\begin{eqnarray*}[X,Y]^{a}=
\{X^a,Y\}',\qquad (\mbox {ad}\, X)\, Y= [X,Y].
\end{eqnarray*}
Then (\ref{qsp}) can be written in the form
\begin {eqnarray}
\label{qsp2}
W K+ [\Omega, \Phi_0]+ A K + [\Pi,K]=0.
\end {eqnarray}
By using (\ref{or7}) we get 
\begin {eqnarray}
\label{qsp3}
K + W^+([\Omega, \Phi_0]+ A K + [\Pi,K])=Y, 
\end {eqnarray}
where 
\begin{eqnarray*}Y\in S^0,\qquad  W{Y}=0,\qquad  \mbox{\rm ngh} ({Y})=0.\end{eqnarray*}

Let us denote $
\bar W=\epsilon_{ab}W^aW^b.
$
Then 
\begin {eqnarray*}
\bar W \bar \Gamma-\bar \Gamma\bar W=4N^2-2MN,
\end {eqnarray*}
from which it follows that for any $X\in {\cal V}$ 
\begin {eqnarray}
\label{qsp5}
X= \frac 1 2 MN^{-1} X + \frac 1 4 \left(
\bar W \bar \Gamma -\bar \Gamma \bar W\right)N^{-2}X.
\end {eqnarray}

The boundary conditions (\ref{pps2}) imply $ \bar \Gamma K=0,$ and therefore $ \bar \Gamma Y=0,$ since $\bar \Gamma W^+=0.$
By using (\ref{qsp5}) we get $Y=0.$

Solving (\ref{qsp3}) for $K$ yields
\begin {eqnarray} 
\label{q21}
K=- (I+W^{+}(A+ {\rm ad}\,\Pi) )^{(-1)}W^+[\Omega, \Phi_0].
\end {eqnarray}  
We must now show that (\ref{q21}) satisfies (\ref{qsp2}).

The Jacobi identities for the functions $\Omega^{\prime a},\Phi' $ imply 
\begin {eqnarray}  
\label {q1}
\{\Omega^{\prime a},\{\Omega^{\prime b},\Phi'\}'\}'+
\{\Omega^{\prime b},\{\Omega^{\prime a},\Phi'\}'\}'=0.
\end {eqnarray} 
Let $R=(R^a)$ denote left-hand side of (\ref{qsp2}), 
\begin {eqnarray}
\label{qsp4}
R=W K+ [\Omega, \Phi_0]+ A K + [\Pi,K]. 
\end {eqnarray}
Then (\ref{q1}) takes the form 
\begin {eqnarray}  
\label {l3}
WR+AR+[\Pi,R]=0.
\end {eqnarray} 
It is easily verified that if $K$ satisfies (\ref{qsp}) then 
${W^{+}K = W^{+}\Upsilon,}$ and
\begin {eqnarray}
\label{s5}
W^{+}R=0.
\end {eqnarray}
We note that $K\in S^0$ and $R\in S^1.$
Consider (\ref{l3}) and (\ref{s5}), where $K$ satisfies (\ref{qsp}). By using (\ref{dor}), we get  
\begin {eqnarray*}
R=- W^{+}( A R +[\Pi,R]). 
\end {eqnarray*} 
From this it follows that $R=0.$ 

We conclude that the solution to Eqs. (\ref{pps}), (\ref{pps2}) is given by  
\begin {eqnarray} 
\label{q2}
\Phi'= L{\Phi_0},
\end {eqnarray}  
where 
\begin {eqnarray*} 
L=I- (I+W^{+}(A+ {\rm ad}\,\Pi) )^{(-1)}W^{+}{\rm ad}\,\Omega.
\end {eqnarray*}  
The operator $L$ is invertible. The inverse $L^{-1}$ is given by 
\begin {eqnarray*} 
L^{-1}\Phi' = \left.\Phi' \right|_{C={\pi} =0}.
\end {eqnarray*}  
Equation (\ref{q2}) establishes a one-to-one correspondence between first class functions and solutions to Eqs. (\ref {pps}), (\ref{pps2}).

Let us denote by $L(A)$ the image of $A\subset P$ under the mapping $L.$  
For $\Phi'_1,\Phi'_2 \in L(P)$ 
\begin {eqnarray*} 
\left.
 \,\,\,\{\Phi'_1,\Phi'_2\}' \right|_{C={\pi} =0}=
\{
\left.\Phi'_1\right|_{C={\pi}=0},
\left.\Phi'_2 \right|_{C={\pi} =0}\}',
\end {eqnarray*}  
\begin {eqnarray*} 
\left.
 (\Phi'_1\Phi'_2) \right|_{C={\pi}=0}=
\left.\Phi'_1\right|_{C={\pi}=0}
\left.\Phi'_2 \right|_{C={\pi}=0}. 
\end {eqnarray*}  
This means that $L(P)$ and ${P}$ are isomorphic as Poisson algebras, and 
therefore  $L(P)/L(J)$ gives a realization of classical observables.

\end{document}